\documentclass[11pt,showpacs]{revtex4-1}

\usepackage{epsfig}
\input epsf.tex
\usepackage{amssymb}
\usepackage{amsmath}
\usepackage{amsthm}
\usepackage{amsopn}

\def\comment#1{}

\def\beq{\begin{equation}}
\def\eeq{\end{equation}}
\def\bea{\begin{eqnarray}}
\def\eea{\end{eqnarray}}

\def\ep{\epsilon\!\!\!/}

\usepackage{ulem}
\usepackage{cancel}
\usepackage{color}

\def\blue#1{\textcolor{blue}{#1}}

\begin{document}

\title{ Laser photons acquire circular polarization by interacting with a Dirac or Majorana neutrino beam}

\author{Rohoollah Mohammadi$^{1}$\footnote{rmohammadi@ipm.ir} and She-Sheng Xue$^{2}$\footnote{xue@icra.it}}
\affiliation{$^{1}$ School of physics, Institute for research in fundamental sciences (IPM), Tehran, Iran.\\
$^{2}$
ICRANet, P.zza della Repubblica 10, I--65122 Pescara, and Physics Department, University of Rome {\it La Sapienza}, P.le Aldo Moro 5, I--00185 Rome, Italy.}

\begin{abstract}
It is shown that for the reason of neutrinos being left-handed and their gauge-couplings being parity-violated, linearly polarized photons acquire their circular polarization by interacting with neutrinos.
Calculating the ratio of linear and circular polarizations of laser photons interacting with either Dirac or Majorana neutrino beam, we obtain this ratio for the Dirac neutrino case, which is about twice less than the ratio for the Majorana neutrino case. Based on this ratio, we discuss the possibility of using advanced laser facilities and the T2K neutrino experiment to measure the circular polarization of laser beams interacting with neutrino beams in ground laboratories. This could be an additional and useful way to gain some insight into the physics of neutrinos, for instance their Dirac or Majorana nature.
\end{abstract}


\maketitle

\section{Introduction}
Since their appearance, neutrinos have always been extremely peculiar. Their
charge neutrality, near masslessness, flavor mixing, oscillation, type of Dirac or Majorana, and in particular parity-violating gauge-coupling have been at the center of a conceptual elaboration and
an intensive experimental analysis that have played a major role in
donating to mankind the beauty of the standard model for particle physics.
Several experiments studying solar, atmospheric and reactor neutrinos in past several years provide strong evidences supporting the existence of neutrino oscillations \cite{osc1}. This implies that neutrinos are not exactly massless although they chirally couple to gauge bosons.
Therefore they cannot
be exactly  two-component Weyl fermions. Instead, it is tempting to think of the nature of neutrinos, they might be fermions of Dirac type or Majorana type \cite{majorana}. Dirac and Majorana neutrinos have different electromagnetic properties  \cite{nieves}, which can be used to determine which type of neutrinos exists. The well-known example is the neutrino-less double beta decay \cite{dbd}, still to be experimentally verified.

Some attention has been recently driven to study the neutrino interactions with laser beams; here we
mention a few examples. The emission of $\nu\bar\nu$ pairs off electrons in a polarized ultra-intense electromagnetic
(e.g., laser) wave field is analyzed in Ref.~\cite{titov}. The electron-positron production rate has been calculated \cite{tinsley} using
neutrinos in an intense laser field. In the frame of the standard
model, by studying the elastic scattering of a muon neutrino on an
electron in the presence of a linearly polarized laser field, multi-
photon processes have been shown \cite{bai}.

In this Letter, we quantitatively show and discuss linearly polarized photons acquire circular polarization by interacting with neutrinos, for the reason that neutrinos are left-handed and possess peculiar couplings to gauge bosons in a parity-violating manner. In particular we quantitatively calculate the circular polarization that a linearly polarized laser beam develops by interacting with a neutrino beam in ground laboratories.
Moreover we show that using advanced laser facility one can possibly measure the circular polarization of laser photons interacting with a Dirac or Majorana neutrino beam produced by neutrino experiments, e.g., the Tokai-to-Kamioka (T2K) \cite{T2K}. This could possibly provide an additional way to study some physics of neutrinos, for example, the nature of neutrinos, Dirac or Majorana type.

We recall that the similar photon-neutrino process was considered for the generation of circular polarizations of Cosmic Microwave Background (CMB) photons \cite{roh} interacting with cosmic background neutrinos (CNB), instead of photons \cite{xue} and electrons in the presence of magnetic field \cite{khodam}. There the main calculation was done to obtain the power spectrum $C_l^V$ of the circular polarization of CMB photons by the forward scattering between CMB photons and CNB neutrinos.

\section{Stokes parameters} The polarization of laser beam is characterized by means of the Stokes parameters:
the total intensity $I$, intensities of linear
polarizations $Q$ and $U$, as well as the intensity of circular
polarizations $V$
indicating
the difference between left- and right- circular polarizations
intensities. The linear polarization can be represented by $P_L\equiv\sqrt{Q^2+U^2}$. An arbitrary polarized
state of a photon $(|k^0|^2=|{\bf k}|^2)$, propagating in the
$\hat z$-direction, is given by \bea
|\epsilon\rangle=a_1\exp(i\theta_1)|\epsilon_1\rangle+a_2\exp(i\theta_2)|\epsilon_2\rangle,\eea
where linear bases $|\epsilon_1\rangle$ and
$|\epsilon_2\rangle$ indicate the polarization states in the $x$-
and $y$-directions, and $\theta_{1,2}$ are initial phases. Quantum-mechanical operators in this
linear bases, corresponding to Stokes parameters, are given by
\bea
\hat{I}&=&|\epsilon_1\rangle\langle\epsilon_1|+|\epsilon_2\rangle\langle\epsilon_2|,\quad
\hat{Q}=|\epsilon_1\rangle\langle\epsilon_1|-|\epsilon_2\rangle\langle\epsilon_2|,\nonumber\\
\hat{U}&=&|\epsilon_1\rangle\langle\epsilon_2|+|\epsilon_2\rangle\langle\epsilon_1|,\quad
\hat{V}=i|\epsilon_2\rangle\langle\epsilon_1|-i|\epsilon_1\rangle\langle\epsilon_2|.
\label{i-v} \eea
An ensemble of photons in a general mixed state
is described by a normalized density matrix $\rho_{ij}\equiv
(\,|\epsilon_i\rangle\langle \epsilon_j|/{\rm tr}\rho)$, and the dimensionless
expectational values for Stokes parameters are given by
\bea
I\equiv\langle  \hat I \rangle &=& {\rm tr}\rho\hat I
=1,\label{i}\\
Q\equiv\langle  \hat Q \rangle &=& {\rm tr}\rho\hat{Q}=\rho_{11}-\rho_{22},\label{q}\\
U\equiv\langle
 \hat U\rangle &=&{\rm tr}\rho\hat{U}=\rho_{12}+\rho_{21},\label{u}\\
V\equiv\langle  \hat V \rangle &=& {\rm
tr}\rho\hat{V}=i\rho_{21}-i\rho_{21}, \label{v}
\eea
where ``$\rm tr$'' indicates the trace in the space of polarization states. As shown below, in a quantum field theory, the density matrix describing Stokes parameters of polarized particle is represented in the phase space $(x,k)$ in addition to the space of polarization states. Eqs.~(\ref{i}-\ref{v}) determine
the relations between Stokes parameters and the
$2\times 2$ density matrix $\rho$ of photon polarization states.


\section{Quantum Boltzmann Equation for Stokes parameters}
We express the laser field strength $F_{\mu\nu}=\partial_\mu A_\nu-\partial_\nu A_\mu$, and 
free gauge field $A_\mu$ in terms of plane wave solutions in the Coulomb gauge \cite{zuber},
 \beq A_\mu(x) = \int \frac{d^3 k}{(2\pi)^3
2 k^0} \left[ a_i(k) \epsilon _{i\mu}(k)
        e^{-ik\cdot x}+ a_i^\dagger (k) \epsilon^* _{i\mu}(k)e^{ik\cdot x}
        \right],\eeq
where $\epsilon _{i\mu}(k)$ are the polarization
four-vectors and the index $i=1,2$, representing two transverse polarizations of a free photon with four-momentum $k$ and $k^0=|{\bf{k}}|$, $k\cdot\epsilon_i=0$ and
$\epsilon_i\cdot\epsilon_j=-\delta_{ij}$. The creation operators
$a_i^\dagger (k)$ and annihilation operators $a_i(k)$ satisfy the canonical commutation relation
\begin{equation}
        \left[  a_i (k), a_j^\dagger (k')\right] = (2\pi )^3 2k^0\delta_{ij}\delta^{(3)}({\bf k} - {\bf k}' ).
\label{comm}
\end{equation}
The density operator describing an ensemble of free photons in the space of energy-momentum and polarization state is given by 
\bea
\hat\rho(x)=\frac{1}{\rm {tr}(\hat \rho)}\int\frac{d^3k}{(2\pi)^3}
\rho_{ij}(x,k)a^\dagger_i(k)a_j(k),
\label{d-m}
\eea
where $\rho_{ij}(x,k)$ is the general density-matrix, analogous to Eqs.~(\ref{i}-\ref{v}), in the space of polarization states with the fixed energy-momentum ``$k$'' and space-time point ``$x$''.
The number operator of photons $
D^0_{ij}(k)\equiv a_i^\dag (k)a_j(k)$ and its
expectational value with respect to the density-matrix (\ref{d-m}) is defined by
\bea
\langle\, D^0_{ij}(k)\,\rangle\equiv {\rm tr}[\hat\rho
D^0_{ij}(k)]=(2\pi)^3 \delta^3(0)(2k^0)\rho_{ij}(x,k).\label{t1}
\eea
The time-evolution of the number operator $D^0_{ij}(k)$ is governed by the Heisenberg equation
\begin{equation}\label{heisen}
   \frac{d}{dt} D^0_{ij}(k)= i[H_I,D^0_{ij}(k)],
\end{equation}
where $H_I$ is the interacting Hamiltonian of photons with other particles in the standard model. Calculating the expectational value of both sides of Eq.~(\ref{heisen}),
one arrives at
the following Quantum Boltzmann Equation (QBE) for the number operator of photons \cite{cosowsky1994},
\bea
(2\pi)^3 \delta^3(0)(2k^0)
\frac{d}{dt}\rho_{ij}(x,k) = i\langle \left[H^0_I
(t);D^0_{ij}(k)\right]\rangle-\frac{1}{2}\int dt\langle
\left[H^0_I(t);\left[H^0_I
(0);D^0_{ij}(k)\right]\right]\label{bo}\rangle,
\eea
where we only consider the interacting Hamiltonian $H^0_I(t)$ of
photons with neutrinos.
The first term on the right-handed side is a
forward scattering term, and the second one is a higher-order collision term, which will be neglected.
In the following, we attempt to calculate the Stokes parameter $V$ of Eq.~(\ref{v}) to show that a linearly polarized laser beam acquires the component of circular polarization while it interacts with a neutrino beam.

\section{Photon-Neutrino interaction}

First we consider Dirac neutrinos $\psi_{\nu_l}$ interacting with charged leptons $\psi_{l}$ and $W^\pm_\mu$ gauge bosons in the standard model \bea\label{lagw}
\pounds_{\rm int} = \frac{g_w}{2\sqrt{2}}\sum_{l=e,\mu,\tau}\left[\bar{\psi}_{\nu_l}\gamma^{\mu}(1-\gamma^5)\psi_{l}W_{\mu}^+\,+\,\bar{\psi}_{l}\gamma^{\mu}(1-\gamma^5)\psi_{\nu_l}W_{\mu}^-\right],
\eea
where we omit the unitary mixing matrix which is not relevant to the following calculations.
Because neutrino masses are approximately zero, compared with their energies ($E_{\nu_l}\gg m_{\nu_l}$), we adopt the two-component theory for zero-mass particles to describe  neutrinos (see for example Ref.~\cite{zuber})
 \beq
 \psi_\nu(x) = \int \frac{d^3 q}{(2\pi)^3}\frac{1}{
\sqrt{2 q^0}} \sum_{r=\pm}\left[ b_r(q) \mathcal{U}_{r}(q)
        e^{-iq\cdot x}+ d_r^\dagger (q) \mathcal{V}_{r}(q)e^{iq\cdot x}
        \right],\label{psi}
\eeq
where neutrino energy $q^0=|\bf q|$, and helicity or chirality states $\mathcal{U}_\pm(q)=-\mathcal{V}_\pm(q)$ are eigenstates of helicity operator ${\bf \Sigma}\cdot{\bf q}/|{\bf q}|$ and chirality operator $\gamma_5$,
($+$) representing the left-handed neutrino $\mathcal{U}_+(q)$ and right-handed anti-neutrino $\mathcal{V}_+(q)$; ($-$) representing the right-handed neutrino $\mathcal{U}_-(q)$ and left-handed anti-neutrino $\mathcal{V}_-(q)$. The annihilation and creation operators
$b_r$ ($d_r$) and  $b^\dagger_r$ ($d^\dagger_r$) for corresponding states satisfy following relations,
\begin{equation}
        \left\{  b_s (q), b_r^\dagger (q')\right\}=\left\{  d_s (q), d_r^\dagger (q')\right\} = (2\pi )^3 \delta_{sr}\delta^{(3)}({\bf q} - {\bf q}' ).
\label{commf}
\end{equation}
For a given momentum $q$ and chirality state ($+$), there exists only two independent states of left-handed neutrino and right-handed anti-neutrino. The chiral projector $(1-\gamma_5)/2$ in the interacting Hamiltonian (\ref{lagw})
implies that only left-handed neutrinos ($+$) participate the weak interaction. The relations
\begin{equation}\label{sn2}
   \bar{\mathcal{U}}_r(q)\gamma^\mu \mathcal{U}_s(q)=2q^\mu\delta_{rs},\,\,\,\,\, \frac{1}{2}\sum_r\bar{\mathcal{U}}_r(q)\gamma^\mu(1-\gamma^5) \mathcal{U}_r(q)=2q^\mu,
\end{equation}
are useful for the following calculations.

\begin{figure}
\begin{center}
  \includegraphics[width=0.6\columnwidth]{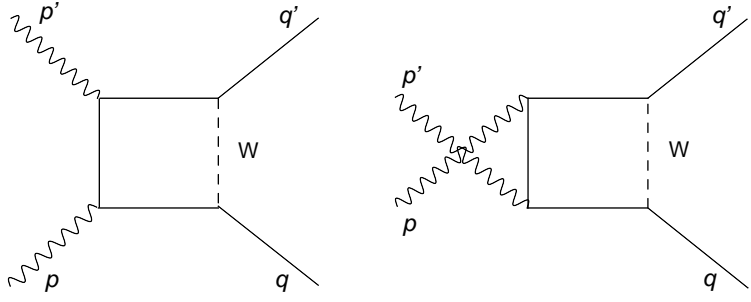}\\
  \caption{Two Feynman diagrams describe the interaction between photons and neutrinos at the one-loop level. $p$ and $p'$ ($q$ and $q'$) represent photons (neutrinos) momenta. Along the loop, there are two QED-vertexes and two vertexes of Eq.~(\ref{lagw}), as well as a $W_\mu^\pm$ gauge-boson propagator $D_{\alpha\beta}$ (dashed line) and three charged lepton propagators $S_F$ (solid lines). Note that the contribution from the neutral channel diagram with a $Z^0_\mu$-boson exchange (see Ref.~\cite{dicus}) is proportional to neutrino masses $m_\nu$, we consider in this Letter exactly massless neutrinos so that the neutral channel diagram with a $Z^0_\mu$-boson exchange does not contribute to the interacting Hamiltonian (\ref{h0}) for studying the generation of circular polarization of laser photons. Even if we consider massive neutrinos, the contribution from the neutral current, being proportional to neutrino masses, can be neglected in comparison with the contribution from charged current.}\label{cmb1}
  \end{center}
\end{figure}
In the context of standard model $SU_{L}(2)\times U_{Y}(1)$ for the electro-weak interactions, two Feynman diagrams representing the leading-order contribution to the interaction between photons and neutrinos are shown in Fig.~\ref{cmb1}.
The leading-order interacting Hamiltonian is given by
\begin{eqnarray}
  H^0_I &=& \int d\mathbf{q} d\mathbf{q'} d\mathbf{p} d\mathbf{p'} (2\pi)^3\delta^3(\mathbf{q'} +\mathbf{p'} -\mathbf{p} -\mathbf{q} ) \nonumber \\
   &\times& \exp[it(q'^0+p'^0-q^0-p^0)]\left[b^\dagger_{r'}a^{\dagger}_{s'}(\mathcal{M}_1+\mathcal{M}_2)a_sb_r\right],\label{h0}
\end{eqnarray}
where
\begin{eqnarray}
  \mathcal{M}_1+\mathcal{M}_2 &=& \frac{1}{8}e^2g_w^2\int\frac{d^4l}{(2\pi)^4}D_{\alpha\beta}(q-l)\bar{\mathcal{U}}_{r'}(q')\gamma^\alpha (1-\gamma_5)S_F(l+p-p')\nonumber\\
   &\times& \left[\ep_{s'}S_F(l+p)\ep_{s}+\ep_{s}S_F(l-p')\ep_{s'}\right]
   S_F(l)\gamma^\beta (1-\gamma_5)\mathcal{U}_r(q),\label{md}
\end{eqnarray}
$D_{\alpha\beta}$ and $S_F$ are respectively $W_\mu^\pm$ gauge-boson and charged-lepton propagators. In Eq.~(\ref{h0}), our notation
$d\mathbf{q}\equiv d^3q/[(2\pi)^32q^0]$, the same for $d\mathbf{p},d\mathbf{p'}$ and $d\mathbf{q'}$.

\section{Laser-photon circular polarization}

From Eq.~(\ref{h0}) for the photon-neutrino interacting Hamiltonian, we obtain the first forward scattering term of Eq.~(\ref{bo})
\begin{eqnarray}
  [H^0_I, D^0_{ij}({\bf k})] &=& \int d\mathbf{q} d\mathbf{q'} d\mathbf{p} d\mathbf{p'} (2\pi)^3\delta^3(\mathbf{q'} +\mathbf{p'} -\mathbf{p} -\mathbf{q} ) (\mathcal{M}_1+\mathcal{M}_2)\nonumber\\
  &\times& (2\pi)^3[b^\dagger_{r'}b_{r}a^\dagger_{s'}a_{s}2p^0\delta_{is}\delta^3({\bf k}-{\bf p})-b^\dagger_{r'}b_{r}a^\dagger_{s'}a_{s}2p'^0\delta_{js'}\delta^3({\bf k}-{\bf p'})].\label{fws}
\end{eqnarray}
Then, using the commutators of Eqs.~(\ref{comm}) and (\ref{commf}), as well as the following expectation values of operators  \cite{cosowsky1994},
\bea \langle \, a_1a_2...b_1b_2...\, \rangle
&=&\langle \,a_1a_2...\, \rangle\langle \,b_1b_2...\, \rangle,
\label{contraction1}\\
\langle \, a^\dag_{s'}(p')a_{s}(p)\, \rangle
&=&2p^0(2\pi)^3\delta^3(\mathbf{p}-\mathbf{p'})\rho_{ss'}(\mathbf{x},\mathbf{q}),
\label{contraction2}\\
\langle \, b^\dag_{r'}(q')b_{r}(q)\, \rangle
&=&(2\pi)^3\delta^3(\mathbf{q}-\mathbf{q'})\delta_{rr'}\frac{1}{2}n_\nu(\mathbf{x},\mathbf{q}),
\label{contraction3}
\eea
where $\rho_{ss'}(\mathbf{x},\mathbf{q})$ is the local matrix density and $n_\nu(\mathbf{x},\mathbf{q})$ is the local spatial density of neutrinos in the momentum state $\mathbf{q}$. Thus we define the neutrino distribution function $n_\nu(\mathbf{x})$ and the average momentum
$\bar{\mathbf{q}}$ of neutrinos as
\begin{equation}\label{nd-n0}
n_\nu(\mathbf{x})=\int \frac{d^3q}{(2\pi)^3} \,\,n_\nu(\mathbf{x},\mathbf{q}),\quad \bar{\mathbf{q}}=\frac{1}{ n_\nu(\mathbf{x})}\int\frac{ d^3q }{(2\pi)^3}\,{\bf q}\,n_\nu(\mathbf{x},\mathbf{q}).
\end{equation}
For a neutrino beam we assume $n_\nu(\mathbf{x},\mathbf{q})\sim \exp[-|\mathbf{q}-\bar{\mathbf{q}}|/|\bar{\mathbf{q}}|]$, representing the most of neutrinos carry the momentum $\mathbf{q}\approx \bar{\mathbf{q}}$.
Then we arrive at
\begin{eqnarray}
  i\langle[H^0_I, D^0_{ij}({\bf k})]\rangle &=& \frac{i}{16}e^2g_w^2\int d\mathbf{q} \big[\rho_{s'j}(\mathbf{x},{\bf k})\delta_{is} -\rho_{is}(\mathbf{x},{\bf k})\delta_{js'}\big]n_\nu(\mathbf{x},\mathbf{q})\nonumber\\
  &\times& \int\frac{d^4l}{(2\pi)^4}D_{\alpha\beta}(q-l)\bar{\mathcal{U}}_{r}(q')\gamma^\alpha (1-\gamma_5)S_F(l)\nonumber\\
   &\times& \left[\,\ep_{s'}S_F(l+k)\,\ep_{s}+\ep_{s}\,S_F(l-k)\ep_{s'}\,\,\right]
   S_F(l)\gamma^\beta (1-\gamma_5)\mathcal{U}_r(q).\label{fws1}
\end{eqnarray}
Moreover, using dimensional regularization and Feynman parameterizations for four-momentum integration over $l$, we approximately obtain
\begin{eqnarray}
  i\langle[H^0_I, D^0_{ij}({\bf k})]\rangle &\approx & -\frac{1}{16}\frac{1}{4\pi^2}e^2g_w^2\int d\mathbf{q} \big[\rho_{s'j}(\mathbf{x},{\bf k})\delta_{is} -\rho_{is}(\mathbf{x},{\bf k})\delta_{js'}\big]n_\nu(\mathbf{x},\mathbf{q})\nonumber\\
  &\times& \int_0^1dy\int_0^{1-y}dz\frac{(1-y-z)}{zM^2_W}\bar{\mathcal{U}}_{r}(q)(1+\gamma_5)\Big[2z q\!\!\!/\epsilon_{s'}\cdot\epsilon_s\nonumber\\
   &+& \,2z(\ep_{s'}\,\,q\cdot\epsilon_s\,+\ep_{s}\,q\cdot\epsilon_{s'}\,)+
  (3y-1)k\!\!\!/\,(\ep_{s}\,\ep_{s'}\,-\ep_{s'}\,\ep_{s}\,)  \Big]\,\mathcal{U}_r(q),
  \label{fws2}
\end{eqnarray}
where neutrino and photon energies are much smaller than $W^\pm_\mu$-boson mass $M_W$, i.e., $E_\nu,\,E_\gamma<<M_W$.

\begin{figure}
  \includegraphics[width=4in]{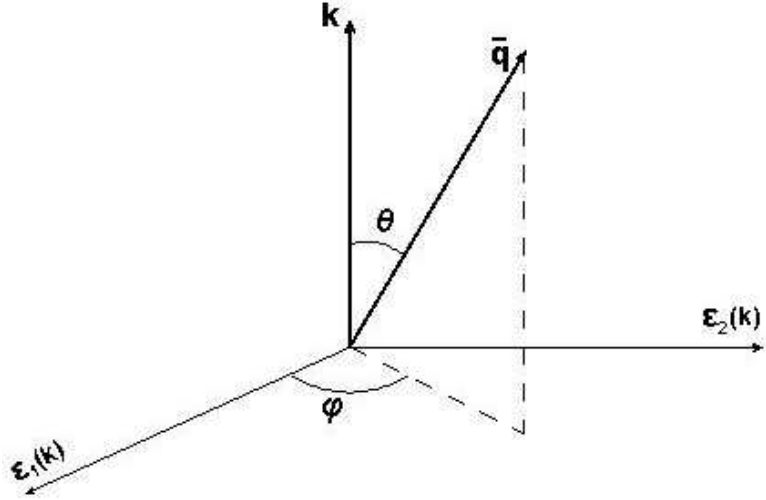}
  \caption{In the laboratory frame, we set the laser-photon momentum $\mathbf{k}$ in the $\hat{z}$-direction (the direction of incident laser beam), polarization vectors $\epsilon_1(\mathbf{k})$ in the $\hat{x}$-direction and $\epsilon_2(\mathbf{k})$ in the $\hat{y}$-direction. This sketch shows the relative angles $\theta$ and $\phi$ between the laser beam direction $\hat{\mathbf{k}}$ and neutrino beam direction $\hat{\bar{\mathbf{q}}}$. }\label{axis}
\end{figure}

The time evolution of Stokes parameter $V$ of linearly polarized laser-photons is determined by Eqs.~(\ref{bo}) and (\ref{fws2}),
\begin{eqnarray}
  \frac{dV(\mathbf{x},\mathbf{k})}{dt}\Big|_{_D} &\approx& \frac{1}{6}\frac{1}{(4\pi)^2}\frac{e^2g_w^2}{M^2_W k^0}\int d\mathbf{q}\,\,n_\nu(\mathbf{x},\mathbf{q})\bar{\mathcal{U}}_{r}(q)(1+\gamma_5) \nonumber\\
  &\times& \left[(\ep_{1}\,q\cdot\epsilon_1-\ep_{2}\,q\cdot\epsilon_2)Q(\mathbf{x},\mathbf{k})-
  (\ep_{1}\,q\cdot\epsilon_2+\ep_{2}\,q\cdot\epsilon_1)U(\mathbf{x},\mathbf{k})\right]\mathcal{U}_r(q),\label{v1}
\end{eqnarray}
where  high-order terms suppressed at least by  $1/M_W^4$ have been neglected.
Using Eq.~(\ref{sn2}), we obtain
\begin{eqnarray}
 \frac{dV(\mathbf{x},\mathbf{k})}{dt}\Big|_{_D}  &\approx& \frac{\sqrt{2}}{3\pi k^0}\alpha\,G_F\int d\mathbf{q}\,\,n_\nu(\mathbf{x},\mathbf{q}) \nonumber\\
  &\times& \left[\,(q\cdot\epsilon_{1}\,\,q\cdot\epsilon_1-q\cdot\epsilon_{2}\,\,q\cdot\epsilon_2)Q(\mathbf{x},\mathbf{k})- (q\cdot\epsilon_{1}\,\,q\cdot\epsilon_2+q\cdot\epsilon_{2}\,\,q\cdot\epsilon_1)U(\mathbf{x},\mathbf{k})\,\right],\label{v2}
\end{eqnarray}
where the Fermi and fine-structure constants are
\begin{equation}\label{cof0}
G_F=\frac{\sqrt{2}}{8}\frac{g_w^2}{M_W^2}\approx1.16\times10^{-5}({\rm GeV})^{-2},\,\,\,\,\alpha=\frac{e^2}{4\pi}=\frac{1}{137}.
\end{equation}
\comment{
Next we can consider $\mathbf{q}$ nearby  $\hat{z}$ direction and  $\mathbf{k}$ and $\mathbf{q}$ is given by
\begin{eqnarray}
  \hat{\mathbf{k}} &=& \sin\theta\cos\phi\, \hat{x}\,+\sin\theta\sin\phi\, \hat{y}+\,\cos \theta \, \hat{z}\label{k}
\end{eqnarray}
Corresponding to $\mathbf{k}$, the polarization state can be written as following
\begin{eqnarray}
  \hat{\epsilon}_1(\mathbf{k}) &=& \cos\theta\cos\phi\, \hat{x}\,+\cos\theta\sin\phi\, \hat{y}-\,\sin \theta \, \hat{z}\label{e1k}\\
   \hat{\epsilon}_2 (\mathbf{k})&=& -\sin\phi\, \hat{x}\,+\cos\phi\, \hat{y}\label{e2k}
\end{eqnarray}}
As follows, we will consider the neutrino beam is produced by high-energy muons ($\mu^\pm$) decay, the beam   angle divergence $\theta_{\rm div} \sim m_{\mu^\pm}/E_{\mu^\pm}\ll 1$, where $E_{\mu^\pm}\,(m_{\mu^\pm})$ is the muon energy (mass). This implies that near to the source of high-energy muons ($\mu^\pm$) decay, the momenta of neutrinos are around the averaged one $\bar{\mathbf{q}}$, given by Eq.(\ref{nd-n0}). In this circumstance, we approximate Eq.~(\ref{v2}) as
\begin{eqnarray}
 \frac{dV(\mathbf{x},\mathbf{k})}{dt}\Big|_{_D}  &\approx& \frac{\sqrt{2}}{6\pi k^0}\alpha\,G_F|\bar{\mathbf{q}}|\,\,n_\nu(\mathbf{x},\bar{\mathbf{q}}) \nonumber\\
  &\times& \left[\,(\hat{\bar{\mathbf{q}}}\cdot\epsilon_{1}\,\,\hat{\bar{\mathbf{q}}}\cdot\epsilon_1-
  \hat{\bar{\mathbf{q}}}\cdot\epsilon_{2}\,\,\hat{\bar{\mathbf{q}}}\cdot\epsilon_2)Q(\mathbf{x},\mathbf{k})-
  (\hat{\bar{\mathbf{q}}}\cdot\epsilon_{1}\,\,\hat{\bar{\mathbf{q}}}\cdot\epsilon_2+
  \hat{\bar{\mathbf{q}}}\cdot\epsilon_{2}\,\,\hat{\bar{\mathbf{q}}}\cdot\epsilon_1)U(\mathbf{x},\mathbf{k})\,\right],\label{v2'}
\end{eqnarray}
where $\hat{\bar{\mathbf{q}}}\equiv \bar{\mathbf{q}}/|\bar{\mathbf{q}}|$ indicates the direction of neutrino beam, and $n_\nu(\mathbf{x})\approx {\rm const.}$ along the beam direction. On the analogy of averaged momentum $\bar{\mathbf{q}}$ of neutrino beam, $\mathbf{k}$ should be understood as an averaged momentum of photons in a laser beam in the direction $\hat{\mathbf{k}}\equiv \mathbf{k}/|\mathbf{k}|$, the Stokes parameters $Q(\mathbf{x},\bf k),U(\mathbf{x},\bf k)$ and $V(\mathbf{x},\bf k)$ are the functions averaged over the momentum distribution of photons in a laser pulse. As shown in Fig.~\ref{axis}, we select the $\hat{\mathbf{k}}$ along the $\hat z$-direction, and $\theta$ and $\phi$ spherical angles of $\hat{\bar{\mathbf{q}}}$ with respect to the $\hat{\mathbf{k}}$ direction. Then Eq.~(\ref{v2'}) becomes
 \begin{eqnarray}
  \frac{dV(\mathbf{x},\bf k)}{dt}\Big|_{_D} &\approx & \frac{\sqrt{2}}{6\pi k^0}\alpha\,G_F\,|\bar{\mathbf{q}}|\,n_\nu(\mathbf{x},\bar{\mathbf{q}})\nonumber\\
  &\times& \left[\,(\sin^2\theta\,\cos2\phi)Q(\mathbf{x},\bf k)-
  (\sin^2\theta\,\sin2\phi)U(\mathbf{x},\bf k)\,\right]\nonumber\\
  &=& \frac{\sqrt{2}}{6\pi k^0}\alpha\,G_F\,|\bar{\mathbf{q}}|\sin^2\theta\,n_\nu(\mathbf{x},\bar{\mathbf{q}}) Q(\mathbf{x},\bf k),\label{v21'}
\end{eqnarray}
where in the third line we set $\phi=0$ in the plane of $\bar{\mathbf{q}}$ and ${\mathbf{k}}$. To discuss possible experimental relevance, we rewrite Eq.~(\ref{v21'}) as
 \begin{eqnarray}
 \frac{\Delta V}{Q} &\equiv& \frac{\Delta V(\mathbf{x},\bf k)}{Q(\mathbf{x},\bf k)}\Big|_{_D}
  \approx \alpha\,G_F\,\left(\frac{\sqrt{2} \,|\bar{\mathbf{q}}|}{6\pi  \,k^0}\right)\sin^2\theta\,n_\nu(\mathbf{x},\bar{\mathbf{q}})\Delta t\nonumber\\
&=& 2.37\cdot 10^{-36} ({\rm cm}^2) \left(\frac{\bar F_\nu(\mathbf{x},\bar{\mathbf{q}})}{k^0}\right)\sin^2\theta\,\Delta t,
  \label{v22'}
\end{eqnarray}
where the averaged energy-flux of neutrino beam
$\bar F_\nu(\mathbf{x},\bar{\mathbf{q}})\equiv c |\bar{\mathbf{q}}|\,n_\nu(\mathbf{x},\bar{\mathbf{q}}).$
Eq.~(\ref{v22'}) represents the ratio of circular and linear polarizations of laser beam interacting with the neutrino beam for a time interval $\Delta t$.


\comment{
In the laboratory frame, setting the laser-photon momentum $\mathbf{k}$ in the $\hat{z}$-direction (the direction of incident laser beam), $ \epsilon_1$ in the $\hat{x}$-direction and $\epsilon_2$ in the $\hat{y}$-direction,
we rewrite Eq.~(\ref{v2}) as
\begin{eqnarray}
  \frac{dV(\bf k)}{dt}\Big|_{_D} &=& \frac{\sqrt{2}}{3\pi k^0}\alpha\,G_F\,\,\frac{1}{2}\int d\Omega_q \varepsilon_\nu(x,\theta,\phi)\nonumber\\
  &\times& \left[\,(\sin^2\theta\,\cos2\phi)Q(\bf k)-
  (\sin^2\theta\,\sin2\phi)U(\bf k)\,\right],\label{v21}
\end{eqnarray}
where the neutrino energy-density is given by
\begin{eqnarray}
 \varepsilon_\nu(x,\theta,\phi)\equiv \int \frac{dq^0}{(2\pi)^3}\, q^0  \,\,n_\nu(q^0,\theta,\phi),\label{eden}
\end{eqnarray}
the neutrino energy $q^0\approx |\bf q|$, the neutrino-number distribution $n_\nu(q_x,q_y,q_z)=n_\nu(q^0,\theta,\phi)$ and the range of angular integration $\int d\Omega_q\equiv \int_0^{2\pi}d\phi\int\sin\theta d\theta $, all of these depend on neutrino beams in experiments. If the neutrino beam is produced by high-energy muons ($\mu^\pm$), $\theta \sim 1/\gamma$ and the Lorentz factor $\gamma=E_{\mu^\pm}/m_{\mu^\pm}$. Note that the neutrino-number distribution cannot be exactly axially symmetric with respect to the momentum $\bf k$ of laser-photons, i.e.~$n_\nu(q^0,\theta,\phi)=n_\nu(q^0,\theta)$, otherwise Eq.~(\ref{v21}) vanishes. In addition, all incoming laser photons cannot strictly carry an identical momentum $\bf k$, instead they have distributions $I(\bf k)$, $Q(\bf k)$ and $U(\bf k)$ which are functions of $\theta$ and $\phi$.  Given the total laser intensity $I(x)=\int d{\bf k}I(\bf k)$, in Eq.~(\ref{v21}), we take an average over the distribution of laser photons, as a result
\begin{eqnarray}
  \frac{dV(x)}{dt}\Big|_{_D} &=& \frac{\sqrt{2}}{3\pi k^0}\alpha\,G_F\,\,\frac{1}{2}\, I(x) \int d\Omega_q \varepsilon_\nu(x,\theta,\phi)\nonumber\\
  &\times& \left[\,(\sin^2\theta\,\cos2\phi)\hat Q(x,\theta,\phi)-
  (\sin^2\theta\,\sin2\phi)\hat U(x,\theta,\phi)\,\right],\label{v22}
\end{eqnarray}
where $x$ indicates the space-time point of a laser beam interacting with a neutrino beam, $\hat Q$ and $\hat U$, represent the normalized components of linear polarizations of the laser beam. In Eq.~(\ref{v22}), we expand the integrand in terms of spherical harmonics
\begin{eqnarray}
\varepsilon_\nu(x,\theta,\phi)
\hat P(x,\theta,\phi)=\sum_{lm}c_{lm}(x)Y_{lm}(\theta,\phi),
\label{h22}
\end{eqnarray}
and $c_{lm}(x)\approx c_{lm}\varepsilon_\nu(x)\hat P(x)$, where $\varepsilon_\nu(x)$ is the neutrino density and $\hat P(x)=\hat Q(x), \hat U(x)$ is the normalized density of linearly polarized photons. As a result, Eq.~(\ref{v22}) becomes
\begin{eqnarray}
  \frac{dV(x)}{dt}\Big|_{_D} &\approx & \frac{\sqrt{2}}{3\pi k^0}\alpha\,G_F\,\,\frac{1}{2}\, I(x) \varepsilon_\nu(x)\left[\,({\rm Re}\,c_{22})\hat Q(x)-
 ({\rm Im}\,c_{22})\hat U(x)\,\right]\nonumber\\
 &= & \frac{\sqrt{2}}{3\pi k^0}\alpha\,G_F\,\,\frac{1}{2}\, I(x) \varepsilon_\nu(x)\,({\rm Re}\,c_{22})\hat Q(x),\label{v33}
\end{eqnarray}
where $c_{22}$ represents the quadrable component of $Y_{22}(\theta,\phi)$, and the second line is given in a particular polarization frame where $\hat U(x)=0$.}

To end this section, we would like to point out that the purely left-handed interaction (\ref{lagw}) is the crucial reason why linearly polarized photons acquire circular polarizations by interacting with left-handed neutrinos, in contrast they do not acquire circular polarizations by interacting with electrons in the forward scattering terms of Eq.~(\ref{bo}) \cite{cosowsky1994}. If right-handed neutrinos were involved in the weak interaction, replacing $(1-\gamma_5)/2$ by $(1+\gamma_5)/2$ in Eq.~(\ref{lagw}), we would have obtained a contribution from right-handed neutrinos that completely cancels the  circular polarization (\ref{v21'}). This can be understood by the angular-momentum conservation in the photon-neutrino interaction (\ref{lagw}), where neutrinos are in left-handed state $(+)$, as a result, photons acquire the component of circular polarizations. This point will be further illustrated by the Compton scattering of photons and polarized electrons in the next section.

\section{Polarized Compton scattering}
\begin{figure}
  \includegraphics[width=4in]{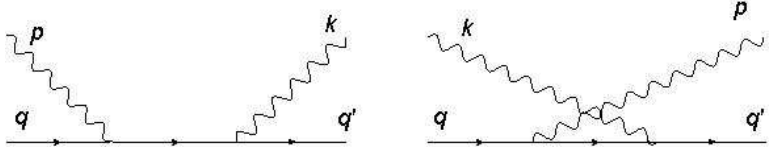}\\
  \caption{The Feynman diagrams for the electron-photon scattering}\label{compton}
\end{figure}

It is shown \cite{cosowsky1994} that the photon circular-polarization is not generated by
the forward Compton scattering of linearly polarized photons on unpolarized electrons. The reason is that the contributions from left- and right-handed polarized electrons to photon circular-polarization exactly cancel each other. In the following, we show the circular polarization generated by a polarized Compton scattering: a photon beam scattering on a polarized electron beam, whose the number-density of incident left-handed electrons is not equal to the right-handed one. For the sake of simplicity, we describe the polarized electron beam by the net number-density of left-handed electrons $\delta n_{L,e}$, and consider relativistic electrons $E_e\gg m_e$,  $E_e\gg E_\gamma$ and $\mathbf{q}\gg\mathbf{k}\sim \mathbf{p}$, where $\mathbf{q}$, $\mathbf{p}$ and $\mathbf{k}$ are the momenta of incident electrons, incident and scattered photons  respectively (see Fig.~\ref{compton}).
Following the calculations of Ref.~\cite{cosowsky1994}, we obtain the time evaluation of the Stokes parameter $V$ for the circular polarization of scattered photons,
\begin{eqnarray}
  \frac{dV(\mathbf{k})}{dt} &\approx& \frac{\delta\bar{n}_{L,e}}{4}\frac{3\sigma_T}{64\pi^2}\left(\frac{m_e}{|\bar{\mathbf{q}}|}\right)^2
  \left(\frac{m_e}{\bar{\mathbf{q}}\cdot\mathbf{k}}\right)^2
  \int\frac{d\Omega_{\hat{\mathbf{p}}}}{4\pi}\nonumber\\
  &\times&\Big\{[\bar{\mathbf{q}}\cdot\epsilon_2(\mathbf{k})]^2-2[\epsilon_1(\hat{\mathbf{p}})\cdot\epsilon_2(\mathbf{k})][\bar{\mathbf{q}}\cdot\epsilon_2(\mathbf{k})]
  [\bar{\mathbf{q}}\cdot\epsilon_1(\hat{\mathbf{p}})]+[\bar{\mathbf{q}}\cdot\epsilon_1(\hat{\mathbf{p}})]^2  -(1\Leftrightarrow 2) \Big\}U(\hat{\mathbf{p}}),\label{compton1}
\end{eqnarray}
which is proportional to $\delta\bar{n}_{L,e}$ the averaged number-density of left-handed polarized electrons, here $\bar{\mathbf{q}}$ is the averaged momenta of incident polarized electrons and $\hat{\mathbf{p}}\equiv \mathbf{p}/|\mathbf{p}|$. This shows that linearly polarized photons scattering on polarized electrons acquire circular polarization, analogously with photon circular-polarization generated by linearly polarized photons scattering on left-handed neutrinos.

\section{Majorana neutrinos}

We turn to calculate the circular polarization of laser-photon beam due to its scattering with Majorana neutrinos \cite{palbook},
\begin{eqnarray}
  \psi^M_\nu(x)
& =& \int \frac{d^3 q}{(2\pi)^3\sqrt{2 q^0}}\sum_{r=\pm}\left[ b_r(q) \mathcal{U}_{r}(q)
        e^{-iq\cdot x}+\lambda\, b_r^\dagger (q) \mathcal{V}_{r}(q)e^{iq\cdot x}
        \right],\label{major}
\end{eqnarray}
which are self-conjugated Dirac neutrinos (particle and anti-particle are identical), i.e., $\psi^M_\nu(x)=\lambda \psi^M_\nu(x)^c$ up to a phase $\lambda$ ($|\lambda^2|=1$), where the conjugated field $\psi^M_\nu(x)^c=\gamma^0 C  \psi^M_\nu(x)^*$, $C=i\gamma^0\gamma^2$, and
\begin{equation}\label{spinor1}   \gamma^0\,C\,\mathcal{V}^*_{r}(p)=\mathcal{U}_{r}(p),\,\,\,\,\,\,\,\,\gamma^0\,C\,\mathcal{U}^*_{r}(p)=\mathcal{V}_{r}(p).
 \end{equation}
The interaction of Majorana neutrinos with charged particles via the $W_{\mu}^\pm$ in the Standard Model $SU_{L}(2)\times U_{Y}(1)$ can be written as \cite{palbook}
\begin{eqnarray}
\pounds^M_{\rm int} = \frac{g_w}{2\sqrt{2}}\sum_{l=e,\mu,\tau}\bar{\psi}^M_{\nu}\left[\gamma^{\mu}(1-\gamma^5)\psi_{l}W_{\mu}^+\,-\,\lambda\gamma^{\mu}(1+\gamma^5)\psi^c_{l}W_{\mu}^-\right],\label{lagwm}
\end{eqnarray}
where $\psi^c_{l}$ is the conjugated lepton field.
The first term is due to the left-handed Majorana neutrino, analogously to Eq.~(\ref{lagw}) for the left-handed Dirac neutrino, yielding Feynman diagrams in Fig.~\ref{cmb1}. The second term comes from the conjugated left-handed Majorana neutrino $\psi^M_\nu(x)^c$, yielding ``conjugated'' Feynman diagrams, which are
all the internal lines in Fig.~\ref{cmb1} are replaced by their conjugated lines. The contribution from
``conjugated''  Feynman diagrams can be obtained by
substitutions \cite{palbook}:
\begin{eqnarray}
  W^+ &\rightarrow& W^-,\,\,\,\,\,\,\,\, \psi_e\rightarrow\psi_e^c,\,\,\,\,\,\,\,
  \gamma^5\rightarrow-\gamma^5, \,\,\,\,\,\,\,\,g_w\rightarrow-\lambda g_w.
\end{eqnarray}
As a result, in Majorana neutrino case, the interacting Hamiltonian is
\begin{eqnarray}
  H^0_I &=& \int d\mathbf{q} d\mathbf{q'} d\mathbf{p} d\mathbf{p'} (2\pi)^3\delta^3(\mathbf{q'} +\mathbf{p'} -\mathbf{p} -\mathbf{q} ) \nonumber \\
   &\times& \exp[it(q'^0+p'^0-q^0-p^0)]\left[b^\dagger_{r'}a^{\dagger}_{s'}(\mathcal{M}_1+\mathcal{M}_2)a_sb_r+
   a^{\dagger}_{s'}b_r(\mathcal{M}'_1+\mathcal{M}'_2)a_sb^{\dagger}_{r'}\right]\nonumber\\
&=& \int d\mathbf{q} d\mathbf{q'} d\mathbf{p} d\mathbf{p'} (2\pi)^3\delta^3(\mathbf{q'} +\mathbf{p'} -\mathbf{p} -\mathbf{q} ) \nonumber \\
   &\times& \exp[it(q'^0+p'^0-q^0-p^0)]\left[b^\dagger_{r'}a^{\dagger}_{s'}(\mathcal{M}_1+\mathcal{M}_2-\mathcal{M}'_1-\mathcal{M}'_2)a_sb_r\right],\label{h0m}
\end{eqnarray}
where $\mathcal{M}_{1}+\mathcal{M}_{2}$ is given by Eq.~(\ref{md}) and the ``conjugated'' part
 \begin{eqnarray}
  -\mathcal{M}'_1-\mathcal{M}'_2 &=& -\frac{1}{8}e^2g_w^2\int\frac{d^4l}{(2\pi)^4}D_{\alpha\beta}(l-q)\bar{\mathcal{V}}_{r}(q)\gamma^\alpha (1+\gamma_5)S_F(p'-p-l)\nonumber\\
   &\times& \left[\ep_{s}\,S_F(-l-p)\,\ep_{s'}\,+\,\ep_{s'}\,S_F(p'-l)\,\ep_{s}\,\right]
   S_F(-l)\gamma^\beta (1+\gamma_5)\mathcal{V}_{r'}(q').
   \label{conj}
\end{eqnarray}
Analogously to Eqs.~(\ref{fws}-\ref{fws2}), we obtain the contribution of ``conjugated'' part (\ref{conj}) to the forward scattering term $ i\langle[H^0_I, D^0_{ij}({\bf k})]\rangle$,
 \begin{eqnarray}
&-&\frac{1}{16}\frac{1}{4\pi^2}e^2g_w^2\int d\mathbf{q}    \big[\rho_{s'j}({\bf x},{\bf k})\delta_{is} -\rho_{is}({\bf x},{\bf k})\delta_{js'}\big]n^M_{\nu}({\bf x},{\bf q})\nonumber\\
  &\times& \int_0^1dy\int_0^{1-y}dz\frac{(1-y-z)}{zM^2_W}\bar{\mathcal{V}}_{r}(q)(1-\gamma_5)\Big[2z q\!\!\!/\epsilon_{s'}\cdot\epsilon_s\nonumber\\
   &+& \,2z(\,\ep_{s'}\,q\cdot\epsilon_s\,+\ep_{s}\,q\cdot\epsilon_{s'}\,)-   (3y-1)k\!\!\!/\,(\ep_{s}\,\ep_{s'}\,-\,\ep_{s'}\,\ep_{s}\,)\,\Big]
  \mathcal{V}_r(q).\label{fws2m}
\end{eqnarray}
As a result, we approximately obtain the time-evolution of the circular-polarization that laser-photons acquire after their interaction with a Majorana neutrino beam,
\begin{eqnarray}
  \frac{dV({\bf x},\mathbf{k})}{dt}\Big|_{_{M}} &\approx& \frac{dV({\bf x},\mathbf{k})}{dt}\Big|_{_{D}} +\frac{1}{6}\frac{1}{(4\pi)^2}\frac{e^2g_w^2}{M^2_W k^0}\int d\mathbf{q}\,\,n^M_\nu({\bf x},{\bf q})\bar{\mathcal{V}}_{r}(q)(1-\gamma_5) \nonumber\\
  &\times& \left[(\ep_{1}\,q\cdot\epsilon_1-\ep_{2}\,q\cdot\epsilon_2)Q({\bf x},\mathbf{k})-  (\ep_{1}\,q\cdot\epsilon_2+\ep_{2}\,q\cdot\epsilon_1)U({\bf x},\mathbf{k})\right]\mathcal{V}_r(q).\label{v1m}
\end{eqnarray}
The first term is the same as Eq.~(\ref{v2}) with the replacement $n_\nu({\bf x},{\bf q})\rightarrow n^M_\nu({\bf x},{\bf q})$, and the rest comes from the so-called ``conjugated''  part, which is the contribution from conjugated left-handed Majorana neutrinos interacting with $W^-_\mu$ in the second term of Eq.~(\ref{lagwm}).
From the identity
\begin{equation}\label{iden}    \bar{\mathcal{V}}_r(q)\gamma^\mu(1\pm\gamma^5)\mathcal{V}_{r'}(q')=\bar{\mathcal{U}}_{r'}(q')\gamma^\mu(1\mp\gamma^5)\mathcal{U}_{r}(q),
\end{equation}
Eq.~(\ref{v1m}) becomes
\begin{eqnarray}
  \frac{dV({\bf x},\mathbf{k})}{dt}\Big|_{_{M}} &\approx& \frac{dV({\bf x},\mathbf{k})}{dt}\Big|_{_{D}} -\frac{1}{6}\frac{1}{(4\pi)^2}\frac{e^2g_w^2}{M^2_W k^0}\int d\mathbf{q}\,\,n^M_\nu({\bf x},{\bf q})\bar{\mathcal{U}}_{r}(q)(1+\gamma_5) \nonumber\\
  &\times& \left[(\ep_{1}\,q\cdot\epsilon_1-\ep_{2}\,q\cdot\epsilon_2)Q({\bf x},\mathbf{k})- (\ep_{1}\,q\cdot\epsilon_2+\ep_{2}\,q\cdot\epsilon_1)U({\bf x},\mathbf{k})\right]\mathcal{U}_r(q).\label{v2m}
\end{eqnarray}
By using identities
\begin{equation}\label{sn2m}
   \bar{\mathcal{U}}_r(q)\gamma^\mu \mathcal{U}_s(q)=2q^\mu\delta^{rs},\,\,\,\,\, \frac{1}{2}\sum_r\bar{\mathcal{U}}_r(q)\gamma^\mu(1\pm\gamma^5)
   \mathcal{U}_r(q)=2q^\mu ,
\end{equation}
we approximately obtain
\begin{eqnarray}
  \frac{dV({\bf x},\mathbf{k})}{dt}\Big|_{_{M}} &\approx& 2\,\frac{dV({\bf x},\mathbf{k})}{dt}\Big|_{_{D}} ,\label{fv2m}
\end{eqnarray}
indicating that the photon circular-polarization generated by the photon-neutrino (Majorana) scattering is about two times larger than that generated by the photon-neutrino (Dirac) scattering.
This difference is due to the fact that in the standard model,
see Eq.~(\ref{lagw}), there is only one species of the left-handed Dirac neutrino field, whereas the left-handed Majorana neutrino field has two species for it being a self-conjugated field, see Eq.~(\ref{lagwm}).
As a result, in the Majorana neutrino case, the interacting Hamiltonian $H_I^0$ of Eq.~(\ref{h0m}) has an additional ``conjugated'' part, compared with the interacting Hamiltonian $H_I^0$ of Eq.~(\ref{h0}) in the Dirac neutrino case.
These two parts in Eq.~(\ref{h0m}) have the same contribution to the $V$-parameter, which is equal to the contribution in the Dirac neutrino case. Thus, the $V$-contribution (\ref{fv2m})
from the photon and Majorana neutrino interaction is twice larger than the $V$-contribution (\ref{v22'}) from photon and Dirac neutrino interaction.

\section{Neutrino and laser-photon beams}
Our result (\ref{v22'}) shows that in order to produce a large intensity of photon circular-polarization for possible measurements, the intensities of neutrino and laser-photon beams should be large enough. In addition the interacting time $\Delta t$ of two beams, i.e., the spatial dimension $\Delta d$ of interacting spot of two beams ($\Delta d\approx c\Delta t$) should not be too small. This depends on the sizes of two beams at interacting point, assuming $\sin^2\theta\sim 1$. Based on the Figs.~ 1 or 25 in Ref.~\cite{T2K} for the Tokai-to-Kamioka (T2K) neutrino experiment (ND280),  the maximal neutrino flux $\sim 10^{13}\,{\rm cm}^{-2}{\rm year}^{-1}$ locating at energies $\bar E_\nu\approx |\bar{\mathbf{q}}|\sim 0.5\, {\rm GeV}$, we estimate the mean energy-flux of muon neutrino beam,
\begin{equation}\label{nflux}
\bar F_\nu(\mathbf{x},\bar{\mathbf{q}})\approx |\bar{\mathbf{q}}|\,n_\nu(\mathbf{x},\bar{\mathbf{q}})c\sim 10^{3}-10^{4} \,{\rm GeV}/({\rm cm}^2{\rm s}).
\end{equation}
The beam angle divergence $\theta_{\rm div} \sim m_\mu/E_\mu \approx 10^{-2}$, where the muon energy $E_\mu\sim \bar E_\nu$.  In addition, in the neutrino experiment (ND280), we estimate the spatial and temporal intervals $\Delta d$  and $\Delta t$ of interacting spot as $\Delta d\sim 2 R_0+d\cdot \theta \sim 100\,{\rm cm}$ and $\Delta t\approx \Delta d/c\sim 10^{-8}\,$ sec., where $R_0$ is the beam radius near to the source of muon neutrinos and $d=2.8\cdot 10^4\,$cm is the distance between the muon neutrino source and interacting spot with laser photons.
Using optic laser beams, laser photon energy $k^0\sim {\rm eV}$, we approximately obtain the ratio of Eq.~(\ref{v22'})
 \begin{eqnarray}
 \frac{\Delta V}{Q} \equiv \frac{\Delta V(\mathbf{x},\bf k)}{Q(\mathbf{x},\bf k)}\Big|_{_D}
  \sim 10^{-23}\sin^2\theta (\Delta t/\rm s.),\label{v23'}
\end{eqnarray}
where the factor $\sin^2\theta$ is the order of unit. Eq.~({\ref{v23'}}) represents the result for laser and neutrino beams (pulses) interacting once only. Suppose that by using mirrors or some laser facilities the laser beam can bent $N$-times in its traveling path, and all laser pulses are identical without interference, so that
each laser pulse can interact with the neutrino beam $N$-times in its path, the ratio of
Eq.~({\ref{v23'}}) is approximately enhanced by a factor of $N$,
 \begin{eqnarray}
 \frac{\Delta V}{Q} \equiv \frac{\Delta V(\mathbf{x},\bf k)}{Q(\mathbf{x},\bf k)}\Big|_{_D}
  \sim 10^{-23}\sin^2\theta\, N\,(\Delta t/\rm s.),\label{v24'}
\end{eqnarray}
since $\sin^2\theta=\sin^2(\pi-\theta)$ for back and forth laser beams in opposite directions. Suppose that mirrors that trap laser beams can reflect $99.999\%$ of laser light
for a narrow range of wavelengths and angles. The absorption coefficient of the mirror is then $C_a\sim 10^{-5}$, and the intensity $Q$ of linearly polarized laser beam is reduced to $Q\rightarrow Q(1-C_a)$ for each reflection, we approximately have $N\approx \sum_{n=0}^N(1-C_a)^n\approx 1/C_a$.
In general, we express our result as
 \begin{eqnarray}
 \frac{\Delta V}{Q} \equiv \frac{\Delta V(\mathbf{x},\bf k)}{Q(\mathbf{x},\bf k)}\Big|_{_D}
 \sim 10^{-26}\left(\frac{\bar F_\nu}{10^{4}{\rm GeV}{\rm cm}^{-2}{\rm s}^{-1}}\right)\left(\frac{\Delta t}{10^{-8}{\rm s}}\right)\left(\frac{10^{-5}}{C_a}\right)\left(\frac{\rm eV}{k^0}\right).\label{v25'}
\end{eqnarray}
Based on the T2K neutrino experiment (ND280) and discussions presented in this Section, to have a significant probability of observing a circularly polarized photon ($\sim {\rm eV}$), the intensity of incident linearly polarized photons of optic laser should at least be:  $Q_{\rm min}\sim 10^{26}\,{\rm eV}\,{{\rm cm}^{-2}{\rm s}^{-1}}\sim 10\,{\rm MW}{\rm cm}^{-2}$, which is not difficult to be achieved in present laser technologies. The larger intensity $Q$ of incident laser beam in linear polarization is, the larger intensity of photons in circular polarizations should be observed. In addition,
a laser or neutrino beam is made by sequent bunches (pulses) of photons or neutrinos, it should be important to synchronize two beams in such a way that the interacting probability of photons and neutrinos bunches is maximized.

For a comparison with Eq.~(\ref{v25'}), using Eq.~(\ref{compton1}) we estimate the quantity $\Delta V/Q$ for a linearly polarized laser beam interacting with a polarized electron beam too. Considering the left-handed polarized electron beam with averaged density $\delta\bar{n}_{L,e}$ and averaged momentum $|\bar{\mathbf{q}}|$, we obtain
\begin{eqnarray}
\frac{\Delta V}{Q} &\sim & 10^{-10} \left(\frac{\bar F_e}{10^{4}{\rm GeV}{\rm cm}^{-2}{\rm s}^{-1}}\right)\left(\frac{\Delta t}{10^{-8}{\rm s}}\right) \left(\frac{10^{-5}}{C_a}\right)
\left(\frac{k^0}{\rm eV}\right)^2,\label{compton2}
\end{eqnarray}
where $k^0$ is the averaged energy of laser photons, $\Delta t$
indicates time interaction, the averaged velocity and energy flux of
the left-handed polarized electron beam $\bar v_e=|\bar{\mathbf{q}}|/m_e <1$ and $\bar F_e\equiv |\bar{\mathbf{q}}|\bar v_e \delta\bar{n}_{L,e}$.

In addition, Ref.~\cite{dicus} discussed the possibility of using intense muon neutrino beams, such as those available at proposed muon colliders, interacting with high powered lasers to probe the neutrino mass. The rate of photon-neutrino scattering ($R_{\gamma\nu}\sim $ 1/year or $3.1\times 10^{-8}/{\rm s}$) was estimated by considering dramatically short pulsed lasers with energies of up to $1.6\times10^7 {\rm J}$ per pulse and very short pulse durations ($\sim {\rm fs}$), which is near to the critical intensity $\sim 10^{28-29}{\rm W}/{\rm cm}^2$ for the production of electron-positron pairs. Even in this extremal powered lasers, the total probability for photon-neutrino scattering is too small to be observed. The main reason is that the interacting rate $R_{\gamma\nu}$ is proportional to the photon-neutrino scattering 
cross-section $\sigma_{\gamma\nu}\propto \alpha^2(G_Fm_\nu)^2$, which is very small.

To compare with the photon-neutrino scattering rate $R_{\gamma\nu}$ obtained in Ref.~\cite{dicus}, we estimate 
the rate of generating circular polarization of laser photons presented in this Letter. The quantity $\Delta V({\bf k})$ of Eq.~(\ref{v22'})
represents the number of circularly polarized photons of energy $|{\bf k}|=k^0\sim {\rm eV}$ per unit area (${\rm cm}^{-2}$), unit time (${\rm s}^{-1}$) in a laser pulse. Therefore the rate of generating circular polarization of laser photons can be estimated by
 \begin{equation}\label{rate}
    R_{_V}\approx \left(\frac{\Delta V}{k^0}\right) \sigma_{\rm laser}\, f_{\rm pulse} \,  \, \tau_{\rm pulse},
\end{equation}
where $\tau_{\rm pulse}$ is the time duration of a laser pulse, the effective area of photon-neutrino interaction is represented by the laser-beam size $\sigma_{\rm laser}$ being smaller than the 
neutrino-beam size $\Delta d$, and the laser repetition rate $f_{\rm pulse}$ is the number of laser pulses per second. To have more efficiency, we assume that laser and neutrino beams are synchronized and the $f_{\rm pulse}$ is equal to the repetition rate of neutrino beam $f_{\rm bunch}$, which is the number of neutrino bunches per second.
In contrast with the rate of photon-neutrino scattering $R_{\gamma\nu}\sim \sigma_{\gamma\nu}\propto \alpha^2(G_Fm_\nu)^2$, the rate $R_{_V}$
of Eq.~(\ref{rate}) linearly depends on $\alpha G_F$ via the $\Delta V({\bf k})$ of Eq.~(\ref{v22'}). This implies that the rate $R_{_V}$
of Eq.~(\ref{rate}) should be much larger than the photon-neutrino scattering rate $R_{\gamma\nu}$ considered in Ref.~\cite{dicus}.

Combining Eqs.~(\ref{v25'}) and (\ref{rate}), we obtain
 \begin{eqnarray}
R_{_V} \simeq
 10^{-26}\left(\frac{\bar F_\nu}{10^{4}{\rm GeV}{\rm cm}^{-2}{\rm s}^{-1}}\right)\left(\frac{\Delta t}{10^{-8}{\rm s}}\right)\left(\frac{10^{-5}}{C_a}\right)\left(\frac{\rm eV}{k^0}\right)\,f_{\rm pulse} \,N_\gamma,\label{rate2}
\end{eqnarray}
where $N_\gamma$ is the number of photons in a laser pulse
\begin{equation}\label{rate3}
    N_\gamma= \frac{Q(k)}{k^0}\,\sigma_{\rm laser}\,\tau_{\rm pulse}= \frac{\varepsilon_{\rm pulse}}{k^0},
\end{equation}
and the total energy of a laser pulse $\varepsilon_{\rm pulse}=Q(k)\sigma_{\rm laser}\,\tau_{\rm pulse}$. The averaged power of linearly polarized laser beam is approximately given by $ \bar{P}_{\rm laser}= f_{\rm pulse}\,\varepsilon_{\rm pulse}$, then Eq.~(\ref{rate2}) becomes
 \begin{eqnarray}
R_{_V}\simeq
 10^{-26}\left(\frac{\bar F_\nu}{10^{4}{\rm GeV}{\rm cm}^{-2}{\rm s}^{-1}}\right)\left(\frac{\Delta t}{10^{-8}{\rm s}}\right)\left(\frac{10^{-5}}{C_a}\right)\left(\frac{\rm eV}{k^0}\right)\, \left(\frac{\bar{P}_{\rm laser}}{k^0}\right)\,\,.\label{rate4}
\end{eqnarray}
As a result, with a neutrino beam $\bar F_\nu\sim 10^{4}\,{\rm GeV}\,{\rm cm}^{-2}\,{\rm s}^{-1}$ and a linearly polarized laser beam of energy $k^0\sim $eV and power $\bar{P}_{\rm laser}\simeq 10$MW, the rate of generating circularly polarized photons $R_{_V} \sim 1/{\rm s}$ ($\sim 3\times10^7/{\rm year}$). This rate should be large enough for observations. 

\comment{And also in order to have $R_{_V}\sim 1/{\rm day}$ ($R_{_V}\sim 1/{\rm year}$), the $\bar{P}\simeq 100W$ ($\bar{P}\simeq 3W$) is needed.
If we use more intense laser beam with energy per pulse $\varepsilon_{_{{\rm pulse}}}\sim 10^{7}{\rm J}$, laser repetition rate
 $f_b\sim 15{\rm Hz}$ (consequently  $\bar{P}\sim 150MW$) and $k^0\sim eV$ such as proposing  in \cite{dicus}, the rate of the generation circular polarized photon reaches about $R_{_V}\sim 15/{\rm s}$ ($\sim 5\times 10^{8}/{\rm year}$).}


\section{ Conclusion and remark.}
We show and discuss the reason why a linearly polarized photon acquires its circular polarization by interacting with a neutrino is due to the fact that the neutrino is left-handed and possesses chiral gauge-coupling to gauge bosons.
Calculating the ratio of linear and circular polarizations of photons interacting with either Dirac or Majorana neutrinos, we obtain that this ratio in the case of Dirac neutrinos is about twice less than the ratio in the case of Majorana neutrinos. Based on quantitative value of this ratio, we discuss the possibility of using advanced laser facilities and the T2K neutrino experiment to measure the circular polarization of laser beam interacting with neutrino beam in ground laboratories.

In Ref.~\cite{dicus}, the rate of neutrino-photon scattering was calculated by the Feynman diagram of one fermion-
loop with exchanging a $Z^0_\mu$-boson (neutral current channel). This rate is very small because the obtained cross-section of
photon-neutrino is the order of $(\alpha G_Fm_\nu)^2$, and
it was concluded that the total probability for photon-neutrino scattering is too small to be observed (having an event rate below 1/year), even at PW laser powers. In our case, the time evaluation of the $V$-parameter (\ref{bo}) for the photon circular polarization can be simply written as
\begin{eqnarray}
\frac{dV}{dt} =[H^0_I,V]+\int dt [H^0_I,[H^0_I,V]],\label{v5}
\end{eqnarray}
where the leading non-trivial contribution comes from the forward scattering amplitude $[H^0_I,V]$, which is the order of $(\alpha G_FQ)$, instead of $(\alpha G_FQ)^2$.
Whereas the second term $\int dt [H^0_I,[H^0_I,V]]$ corresponds to the high-order contributions. This implies that our results (\ref{fv2m}) and (\ref{v25'}) of generating circular polarization of laser beam by interacting with neutrino beam might be experimentally measurable.
This could be an additional and useful way to gain some insight into the physics of neutrinos, for instance their Dirac or Majorana nature, moreover their mixing with the right-handed sterile neutrino \cite{mx2014}. 

\section*{\small Acknowledgment}
We thanks the anonymous referee for his/her comments and suggestions.



\begin{thebibliography}{99}
\bibitem{osc1}  Particle Data Group, The European Physical Journal C - Particles and Fields, Volume {\bf3}, Issue 1-4, pp. 1-783, (1998).

\bibitem{majorana}
E.~Majorana, 
Nuovo.\ Cim.\ {\bf14}, 171184 (1937).

\bibitem{nieves}
J. F. Nieves, Phys.\ Rew.\  {\bf D26}, 3152 (1982).

\bibitem{dbd}
V.~Tello, M.~Nemevsek, F.~Nesti and G.~Senjanovic, 
Phys.\ Rev.\ Lett.\ {\bf106}, 151801 (2011);\\
J.~Schechter, J.~W.~F.~Valle, 
Phys.\ Rev.\ {\bf D25}, 2951 (1982).

\bibitem{titov} A.I. Titov, B. Kampfere, H. Takabe and A. Hosaka, Phys.\ Rev.\ {\bf D83}, 053008 (2011).

\bibitem{tinsley}T.~M.~Tinsley,
  Phys.\ Rev.\   {\bf D71}, 073010 (2005),
  [hep-ph/0412014].

\bibitem{bai} Liang Bai, Ming-Yang Zheng and Bing-Hong Wang, Phys.\ Rev.\ {\bf A85}, 013402 (2012).

\bibitem{T2K}
T2K collaboration, Phys.\ Rev.\ {\bf D87}, 012001 (2013);\\
see also,  S.~Geer,
Phys.\ Rev.\ {\bf D57}, 6989 (1998)  [Erratum-ibid.\ D  59, 039903 (1999)].

\bibitem{roh} R. Mohammadi, [arXiv:astro-ph.CO/1312.2199v1].

\bibitem{xue} I.~Motie and S.~-S.~Xue,
Europhys.\ Lett.\  {\bf 100}, 17006 (2012), [arXiv:hep-ph/1104.3555].

\bibitem{khodam}  M.~Giovannini and K.~E.~Kunze, Phys.\ Rev.\ {\bf D78}, 023010 (2008), [arXiv:astro-ph/0804.3380];\\ M.~Giovannini, (2002), [arXiv:hep-ph/0208152];\\
M.~Zarei, E.~Bavarsad, M.~Haghighat, R.~Mohammadi, I.~Motie, Z.~Rezaei, Phys.\ Rev.\ {\bf D81}, 084035 (2010), [arXiv:hep-th/0912.2993].

\bibitem{zuber}
C.~Itzykson and J.~B.~Zuber: ``{\it Quantum field theory}'', McGraw-Hill (1980).

\bibitem{cosowsky1994}
A.~Kosowsky, Annals Phys.\ {\bf246}, 49-85 (1996),
[arXiv:astro-ph/9501045].

\bibitem{dicus}
D.~A.~Dicus, W.~W.~Repko and R.~Vega,
Phys.\ Rev.\ {\bf D62}, 093027 (2000).

\bibitem{palbook}
Rabindra N. Mohapatra and Palash
B. Pal, ``{\it Massive Neutrinos in Physics and Astrophysics}'', Third
Edition, World Scientific (2004).

\bibitem{mx2014}
Rohollah Mohammadi and She-Sheng Xue, in preparation.

\end{thebibliography}
\end{document}